\documentclass{PoS}
\usepackage{amsmath}

\newcommand{\colorful}[0]{CoLoRFulNNLO}

\title{CoLoRFulNNLO for LHC processes}

\ShortTitle{CoLoRFulNNLO}

\author{\speaker{Adam Kardos}\thanks{A.K. acknowledges financial support from
the Premium Postdoctoral Fellowship program of the Hungarian  Academy  of Sciences.
This work was supported by grant K 125105 of the National Research, Development and
Innovation Fund in Hungary.}\\
        Institute of Physics, University of Debrecen\\
        E-mail: \email{kardos.adam@science.unideb.hu}}

\author{Giuseppe Bevilacqua\\
        MTA-DE Particle Physics Research Group, University of Debrecen\\
        E-mail: \email{giuseppe.bevilacqua@science.unideb.hu}}   

\author{G\'abor Somogyi\\
        MTA-DE Particle Physics Research Group, University of Debrecen\\
        E-mail: \email{somogyi.gabor@science.unideb.hu}}

\author{Zolt\'an Tr\'ocs\'anyi\\
        Institute of Physics and MTA-DE Particle Physics Research Group, University of Debrecen\\
        E-mail: \email{zoltan.trocsanyi@cern.ch}}

\author{Zolt\'an Tulip\'ant\\
        Institute of Physics, University of Debrecen\\
        E-mail: \email{tulipant.zoltan@science.unideb.hu}}

\abstract{In my talk I gave a status update on the extension of the CoLoRFulNNLO subtraction method for
computing QCD jet cross sections with hadrons in the initial state. The scheme has been fully worked out
previously for electron-positron collisions and recently important steps have been made towards generalizing 
it to be able to deliver corrections of the same order for LHC processes as well. In particular, the 
important bottleneck of regularizing multiple real emissions has been addressed. We demonstrate the 
numerical stability of the CoLoRFulNNLO method by computing the doubly real contribution for Higgs-boson 
production in gluon-gluon fusion and for W production.}

\FullConference{Loops and Legs in Quantum Field Theory (LL2018)\\
		29 April 2018 - 04 May 2018\\
		St. Goar, Germany}

\begin{document}

\section{Introduction}

As data taking continues at LHC there is an increasing interest in high precision predictions needed for 
interpreting these data. In recent years several advances were made in computing multi-loop amplitudes, 
involving both new techniques and tools. These amplitudes are important ingredients of higher-order 
computations, hence they are essential in making precise predictions. These developments pave the road 
towards automated NNLO calculations and beyond.

Besides the calculation of multi-loop amplitudes, another major issue of higher-order calculations 
must be addressed, namely the regularization of multiple emissions. When extra massless particles appear in 
the final state at higher orders, collinear or soft configurations result in kinematic singularities which 
must be regularized. Several different regularizations are possible: the region of phase space close to the 
kinematic singularity can be cut out (slicing) \cite{Catani:2007vq,Boughezal:2015dva,Gaunt:2015pea} or the 
kinematic behavior of the contribution can be mimicked near the singular limits and can be subtracted 
(subtraction) \cite{Somogyi:2006da,Somogyi:2006db,GehrmannDeRidder:2007jk,Gehrmann:2011wi,GehrmannDeRidder:2012ja,Czakon:2014oma,Caola:2017dug,Magnea:2018hab}. Either way, in order to get a physical result the sliced away or 
subtracted contributions have to be added back after summing and integrating over the unresolved degrees of 
freedom. The CoLoRFulNNLO method was developed as a subtraction scheme using completely local counterterms 
derived from the infrared factorization properties of QCD squared matrix elements. To date, it is worked out 
in full detail for electron-positron annihilation. It was used to obtain predictions for standard event 
shapes with unprecedented numerical precision and to compute observables for which predictions at NNLO 
accuracy had not been available in the literature \cite{DelDuca:2016csb,DelDuca:2016ily,Tulipant:2017ybb}. 
However, in order to handle LHC processes, the method must be extended to be applicable when colored 
particles are present in the initial state. In this proceedings a status update is given on extending the 
scheme and its numerical implementation to LHC processes. In particular, the correct regularization of doubly real emission is 
demonstrated for Higgs-boson production in gluon-gluon fusion and W-boson production.

\section{The Method}

For an infrared-safe observable $J$ computed in some hadron-initiated process the cross section
takes the form of:
\begin{align}
\sigma[J](p_{A},p_{B}) &=
\sum_{a,b}\int_{0}^{1}{\rm d}x_{a}\,{\rm d}x_{b}	f_{a/A}(x_{a};\mu_F^2)f_{b/B}(x_b;\mu_F^2)
\sigma_{ab}[J](p_{a},p_{b};\mu_F^2)
\,.
\end{align}
Here $p_{A}$ and $p_{B}$ are the momenta for the incoming hadrons, $x_{a}$ and $x_{b}$ are 
Bjorken's $x$s, $f_{a/A}$ and $f_{b/B}$ are the parton distribution functions (PDFs), $\mu_F$ is the
factorization scale and $\sigma_{ab}$ is the partonic cross section defined with partons $a$ and 
$b$ as incoming with their momenta obtained as $p_a = x_a p_A$ and $p_b = x_b p_B$.

The partonic cross section has the usual perturbative expansion:
\begin{align}
\sigma_{ab}[J] &= \sigma^{\rm LO}_{ab}[J] + \sigma^{\rm NLO}_{ab}[J] + \sigma^{\rm NNLO}_{ab}[J] 
+\dots
\end{align}
where $\sigma^{\rm LO}_{ab}[J]$, $\sigma^{\rm NLO}_{ab}[J]$ and $\sigma^{\rm NNLO}_{ab}[J]$ represent the
contributions at increasing orders in terms of $\alpha_{\rm S}$. If the lowest
order contribution has $m$ partons in the final state the first term in the expansion can be written as:
\begin{align}
\sigma^{\rm LO}_{ab}[J] &= \int_{m}{\rm d}\sigma_{ab}^{\rm B}J_{m}
\end{align}
where ${\rm d}\sigma^{\rm B}_{ab}$ is the fully differential Born cross section with $a$ and $b$
as initial state partons.

The NLO contribution can be written as the sum of two terms:
\begin{align}
\sigma^{\rm NLO}_{ab}[J] &=
\int_{m+1}{\rm d}\sigma_{ab}^{\rm R}J_{m+1} +
\int_{m}\left\{
{\rm d}\sigma_{ab}^{\rm V} + {\rm d}\sigma_{ab}^{\rm C}
\right\}J_{m}
\,.
\end{align}
Here ${\rm d}\sigma_{ab}^{\rm R}$ stands for the real-emission contribution, 
${\rm d}\sigma_{ab}^{\rm V}$ is the one-loop term and ${\rm d}\sigma_{ab}^{\rm C}$ is
the collinear counterterm. In the \colorful\ method local subtraction terms are used to
regularize kinematic singularities arising from unresolved real emission. These
subtraction terms are derived from the infrared factorization properties of QCD 
amplitudes. With subtractions the NLO contribution takes the form of:
\begin{align}
\sigma^{\rm NLO}_{ab}[J] &=
\int_{m+1}
\left[
{\rm d}\sigma_{ab}^{\rm R}J_{m+1}
- {\rm d}\sigma_{ab}^{\rm R,\,A_1}J_{m}
\right]_{d=4} +
\int_{m}
\left[
\left\{
{\rm d}\sigma_{ab}^{\rm V} + \int_{1}{\rm d}\sigma_{ab}^{\rm R,\,A_1}
+ {\rm d}\sigma_{ab}^{\rm C}
\right\}
J_{m}
\right]_{d=4}
\,.
\end{align} 

At NNLO the contribution is composed of three different terms:
\begin{align}
\sigma^{\rm NNLO}_{ab}[J] &=
\sigma^{\rm NNLO}_{ab,\,m+2}[J] +
\sigma^{\rm NNLO}_{ab,\,m+1}[J] +
\sigma^{\rm NNLO}_{ab,\,m}[J] =
\nonumber\\
&=
\int_{m+2}{\rm d}\sigma_{ab}^{\rm RR}J_{m+2} +
\int_{m+1}\left\{
{\rm d}\sigma_{ab}^{\rm RV} + {\rm d}\sigma_{ab}^{\rm C_1}
\right\}J_{m+1} +
\int_{m}\left\{
{\rm d}\sigma_{ab}^{\rm VV} + {\rm d}\sigma_{ab}^{\rm C_2}
\right\}J_{m}
\end{align}
where ${\rm d}\sigma_{ab}^{\rm RR}$, ${\rm d}\sigma_{ab}^{\rm RV}$ and 
${\rm d}\sigma_{ab}^{\rm VV}$ are the double-real, real-virtual and double-virtual contributions. These 
contain two extra real partons, one extra real parton with one more
loop and two extra loops as compared to the Born process. ${\rm d}\sigma_{ab}^{\rm C_1}$ and 
${\rm d}\sigma_{ab}^{\rm C_2}$ are the collinear counterterms.
As in the case of the NLO correction we define local subtraction terms to regularize
kinematic singularities in the $m+2$ and $m+1$ parton contributions. In the double-real piece up to two 
partons can become unresolved. This is mirrored by the structure of our subtractions:
\begin{align}
\sigma^{\rm NNLO}_{ab,\,m+2}[J] &= 
\int_{m+2}
\left[
{\rm d}\sigma_{ab}^{\rm RR}J_{m+2}
- {\rm d}\sigma_{ab}^{\rm RR,A_2}J_{m}
- {\rm d}\sigma_{ab}^{\rm RR,A_1}J_{m+1}
+ {\rm d}\sigma_{ab}^{\rm RR,A_{12}}J_{m}
\right]_{d=4}
\end{align}
where the last term is introduced to remove the overlap between singly ($A_1$)  and doubly ($A_2$) 
unresolved subtractions.
As for the real-virtual piece the structure is very similar to the one we have already
encountered at NLO due to the presence of only one extra parton. The only
difference is the presence of the integrated $A_1$ terms
necessitating a further subtraction:
\begin{align}
\sigma^{\rm NNLO}_{ab,\,m+1}[J] &= 
\!\!\!\!\int_{m+1}\!\!\!\!\left[
\left\{
{\rm d}\sigma_{ab}^{\rm RV}
\!\!+ \!\!\!\int_{1}\!\!{\rm d}\sigma_{ab}^{\rm RR,A_1}
+ {\rm d}\sigma_{ab}^{\rm C_1}
\right\}J_{m+1}
\!\!-\!\!
\left\{
{\rm d}\sigma_{ab}^{\rm RV,A_1}
\!\!+\!\! \left(
\int_{1}{\rm d}\sigma_{ab}^{\rm RR,A_1}
\right)^{\!\!\!\!\rm A_1}\!\!
\right\}J_{m}
\right]_{d=4}
\!\!\!\!.
\end{align}
Above ${\rm d}\sigma_{ab}^{\rm RV,A_1}$ is understood to include subtraction terms regularizing the real-virtual and the $C_1$ counterterm as well.
The $m$ parton contribution is free from kinematic singularities due to the infrared finiteness of
observable $J$ but because of the two extra loops it contains explicit $\epsilon$ poles. These
poles are cancelled by the integrated forms of the various subtraction terms:
\begin{align}
\sigma^{\rm NNLO}_{ab,\,m}[J] &=
\!\!\!\!\int_{m}\!\!\left[
{\rm d}\sigma_{ab}^{\rm VV} \!\!+ {\rm d}\sigma_{ab}^{\rm C_2}
\!+\!\!\int_{2}\left[
{\rm d}\sigma_{ab}^{\rm RR,A_{2}} \!\!- {\rm d}\sigma_{ab}^{\rm RR,A_{12}}
\right]
\!\!+\!\!\!\!\int_{1}\!\!\left\{
{\rm d}\sigma_{ab}^{\rm RV,A_{1}}
\!\!+\!\! \left(
\int_{1}{\rm d}\sigma_{ab}^{\rm RR,A_{1}}
\!\!\right)^{\!\!\!\rm A_{1}}\!\!\!
\right\}\!
\right]_{d=4}\!\!\!\!\!\!\!\!J_{m}
\,.
\end{align}

The cancelation of the $\epsilon$ poles coming from loop
contributions against those of the integrated subtractions can serve as a powerful check of the subtractions 
and the method. However, the  
drawback of this check is that it requires all the subtractions to be defined and to carry out
the integrations over the unresolved degrees of freedom. Nonetheless, the subtractions
applied in the $m+2$ and $m+1$ parton pieces can be checked separately even before carrying out the
tedious integrations. In the following we summarize those checks we applied to the $m+2$ parton 
contribution in order to validate our subtraction terms.

\section{Checking the Subtraction Terms}

\begin{figure}[!t]
\centering
\begin{tabular}{cc}	
\includegraphics[width=0.5\textwidth]{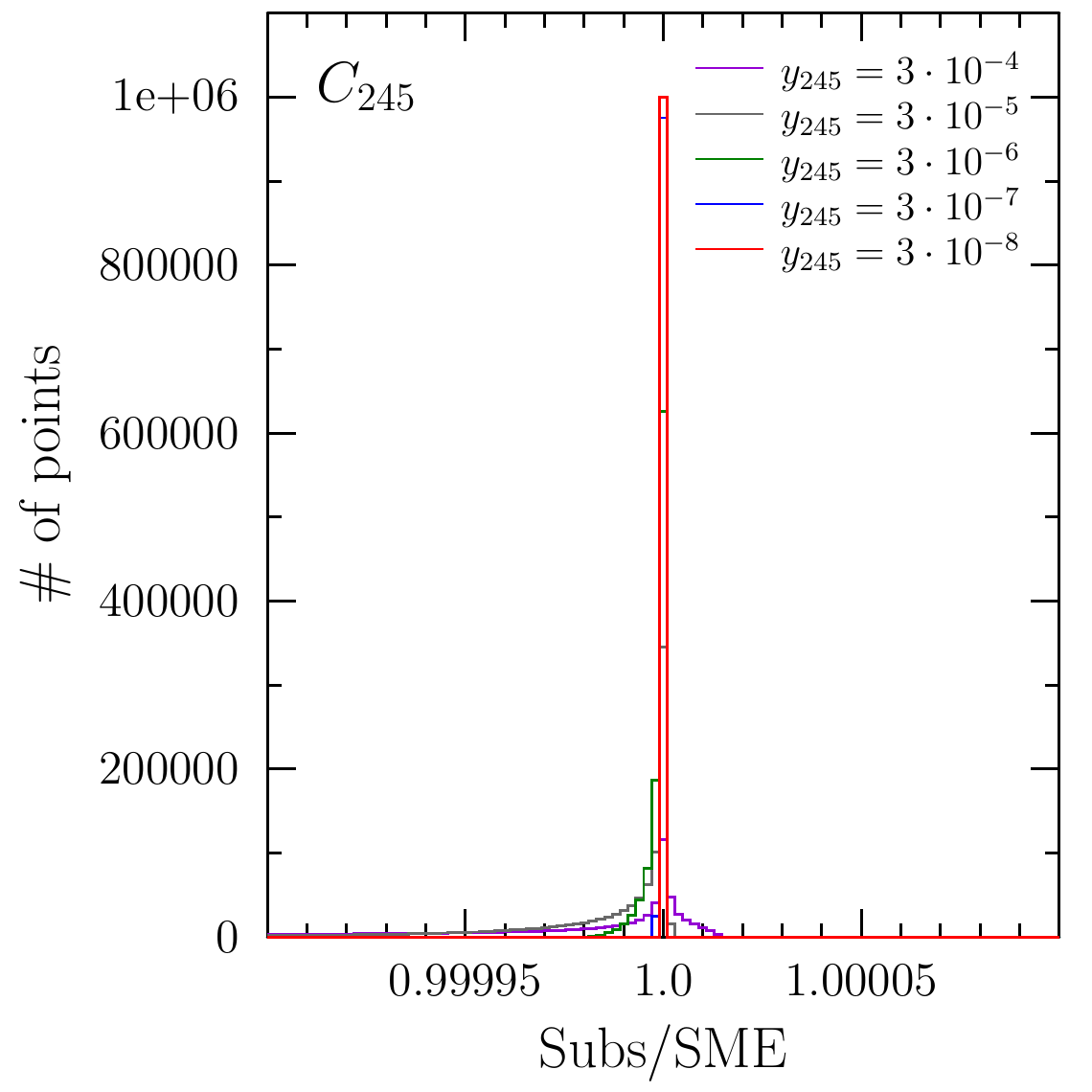} &
\includegraphics[width=0.5\textwidth]{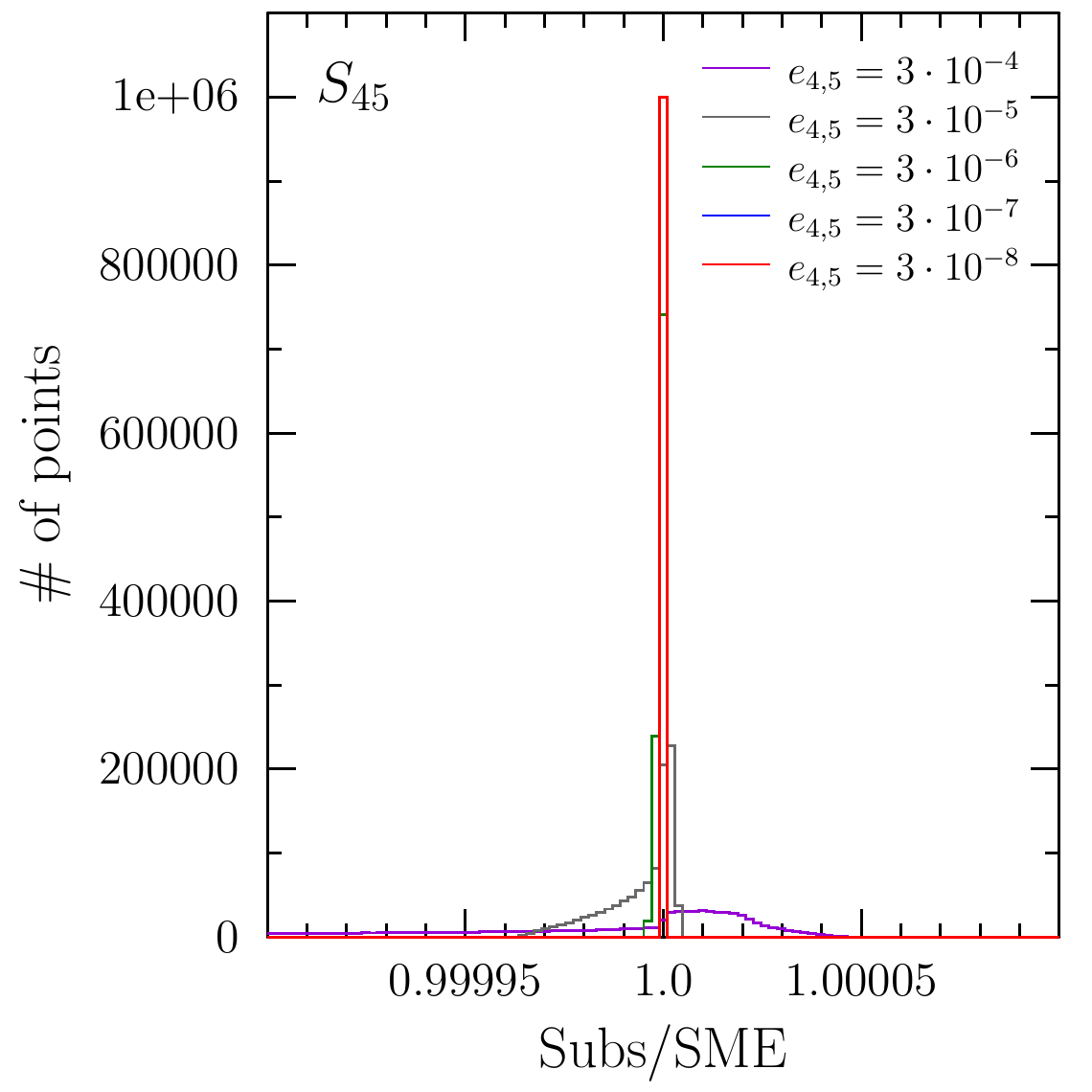}
\end{tabular}
\caption{{\label{fig:spikes}} The rate of cancellation  of kinematic singularities for the process of $u(p_1)+\bar{d}(p_2)\to W^+(p_3)+g(p_4)+g(p_5)$ for a triple 
collinear ($C_{245}$) and a double soft ($S_{45}$) limit. The scaleless quantities 
describing the limit are the scaleless three-particle invariant, $y_{245} = p_{245}^2/Q^2$, and scaleless energies, $e_i = E_i/Q$, respectively.}
\end{figure}

If for a given contribution all subtraction terms are defined and a physical limit is 
approached one subtraction term should cancel the kinematic singularity coming from the squared 
matrix element while the other subtractions should cancel among each other.
This provides means for testing the method. Phase space points approaching any specific 
unresolved configuration can be generated and
the ratio of the sum of all subtractions and the squared matrix element can be computed. If the 
number of points is plotted as a function of the corresponding ratio as the limit is approached the 
spread of points should decrease and ultimately close to the limit all the points should scatter 
around one. We illustrate this for a triple collinear and a double soft limit on Fig. \ref{fig:spikes} where
we present such spike plots in 
case of W production. We checked that all the other limits behaved in a similar way. This provides a 
local check of the subtraction terms.

Furthermore a global test of convergence can be performed.
In any computation beyond leading order accuracy there is a technical cut\footnote{This cut is not to be confused with the cut used throughout in computations performed with the slicing method.} on the phase space 
due to the finite precision of numbers in computer arithmetics. In practice this means that there
is a lower limit defined for all two-particle invariants such that
\begin{align}
y_{\min} &= \min_{i,\,j}\frac{(p_i + p_j)^2}{Q^2} \ge y_{\rm cut}
\,,
\end{align}
where $Q$ is the center-of-mass energy of incoming partons and $y_{\rm cut}$ should be selected small enough such 
that the cross section
is unaffected by further decreasing its value. In case of initial state hadrons those invariants also 
have to be incorporated in the set of pairs that are formed with the initial state partons.

If kinematic singularities are regularized in the double-real and real-virtual contributions and the
resulting cross section contribution is plotted as a function of $y_{\rm cut}$ at small values 
a saturation of the result has to occur. If this does not happen at least one kinematic 
singularity is not
regularized adequately. Plotting the cross section as a function $y_{\rm cut}$ can thus act as a very
powerful check even in the early stages of developing a new subtraction scheme since it only
requires the definition of subtractions for the double-real or real-virtual contribution but not 
their integrated forms.

The \colorful\ framework was first developed for electron-positron annihilation, hence, its
extension to hadronic initial states requires that several new subtraction terms are introduced. 
As the subtraction 
terms are defined on the whole phase space, cancellations must happen between them in various 
limits, which can only be achieved if the Sudakov parameterizations (e.g.: explicit definitions of 
momentum fractions appearing in the splitting kernels) used in the terms are delicately tuned 
to each other. Thus it was essential to perform saturation tests for all the subprocesses of a 
given process in order to ensure correct behavior even if all subtraction terms showed correct limiting 
behavior in their respective limits.

\begin{figure}[!t]
\centering
\includegraphics[width=12cm]{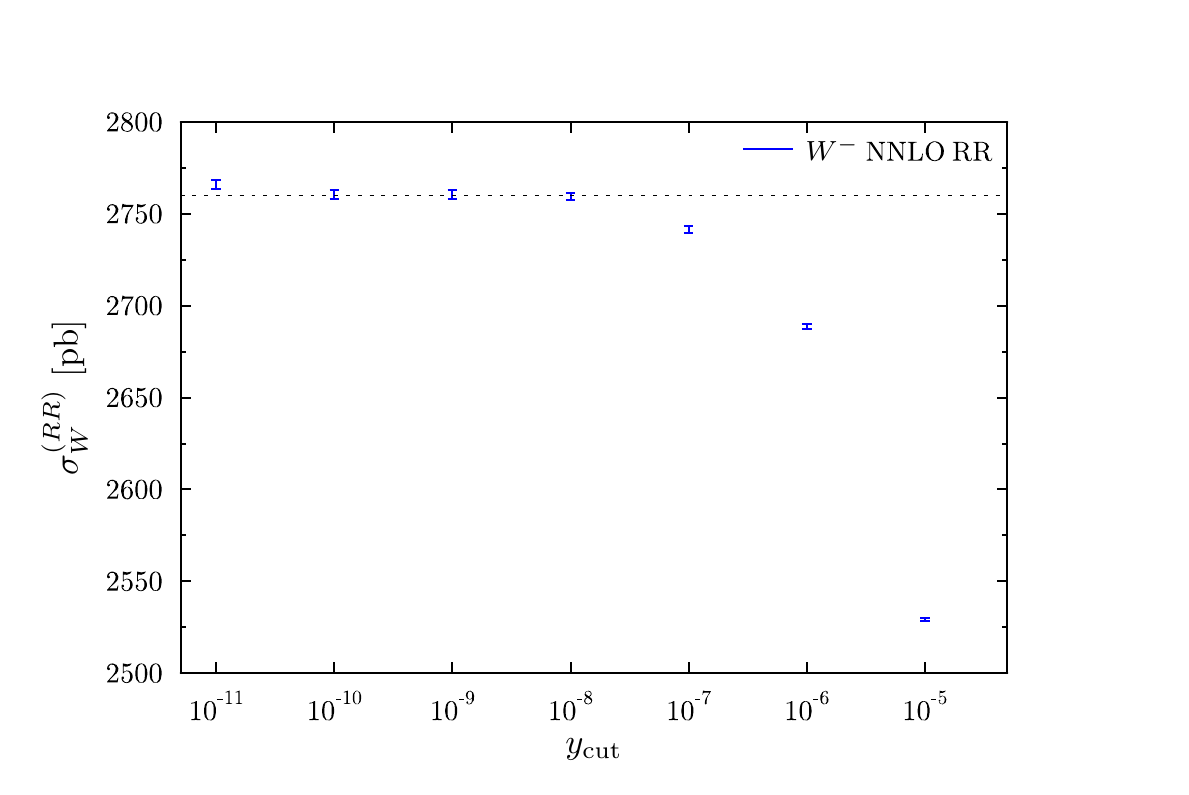}
\caption{{\label{fig:wmsatplot}}The cross section contribution from the double-real line as a function
of $y_{\rm cut}$ for $W^-$ production.}
\end{figure}

\begin{figure}[!t]
\centering
\includegraphics[width=12cm]{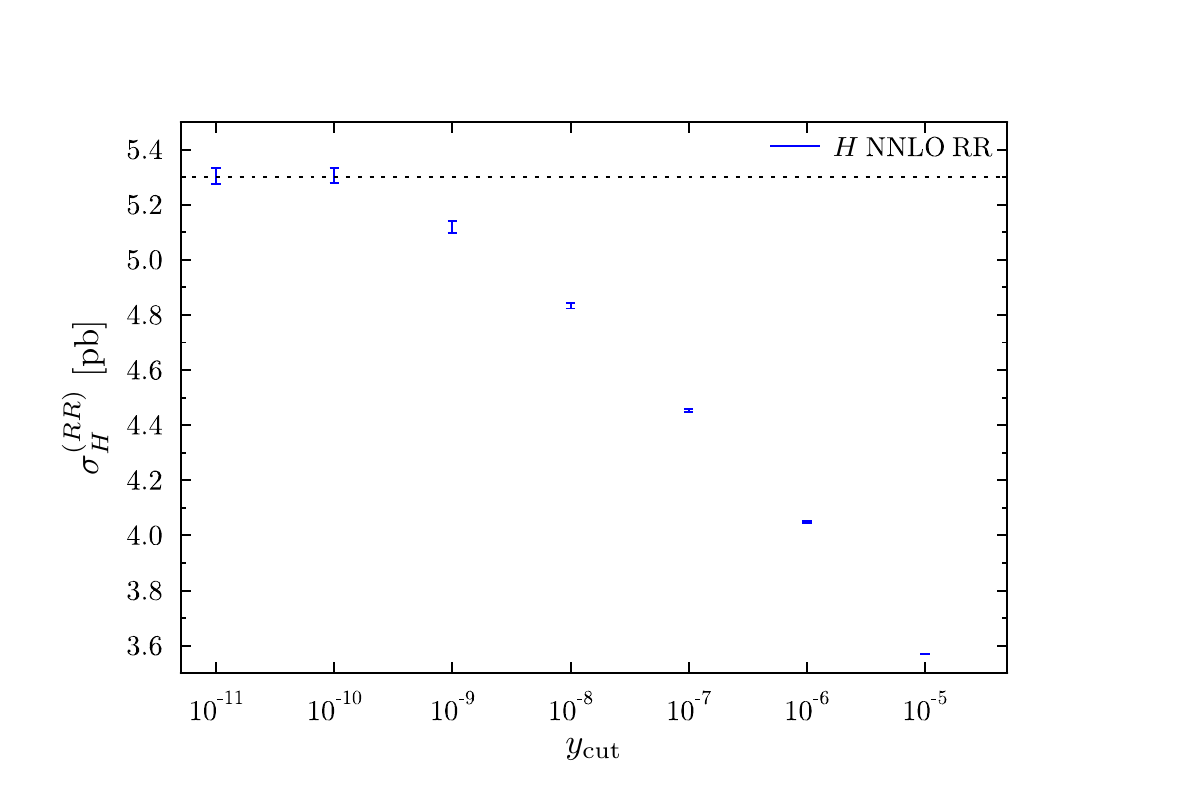}
\caption{{\label{fig:hsatplot}}Same as Fig. \ref{fig:wmsatplot} but for Higgs-boson production.}
\end{figure}

\section{Preliminary Results}

The \colorful\ subtraction method was implemented in the \texttt{MCCSM} (Monte Carlo for the \colorful\ 
Subtraction Method) numerical code \cite{Kardos:2016pic} that is being extended to 
treat processes with hadronic initial states as well. The first processes we implemented are $W^\pm$ 
production and Higgs-boson production in gluon-gluon fusion. For both processes we computed
the cross section contribution coming from the regularized double-real line as a function of $y_{\rm cut}$.

The cross section is unphysical due to the missing real-virtual and double-virtual pieces.
The setup employed in the runs was the following: a $13\,{\rm TeV}$ LHC configuration was used with 
the \verb+NNPDF30_nnlo_as_0118+ PDF set as provided by \texttt{LHAPDF} \cite{Buckley:2014ana}, for
$W^\pm$ production $m_W = 80.385\,{\rm GeV}$, $m_Z = 91.1876\, {\rm GeV}$ and $\alpha_{\rm EM}(m_Z) = 1/128$
and for Higgs-boson production $m_H = 125\,{\rm GeV}$ and vacuum expectation value $v = 246.219\,{\rm GeV}$ were used.

The saturation plot for $W^-$ production is depicted on Fig. \ref{fig:wmsatplot} while for Higgs-boson
production on Fig. \ref{fig:hsatplot}. The saturation of the cross section contribution at small values of 
$y_{\rm cut}$ is visible for both processes indicating that our subtraction terms cancel all 
kinematic singularities correctly.

\begin{figure}[!t]
\centering
\includegraphics[width=12cm]{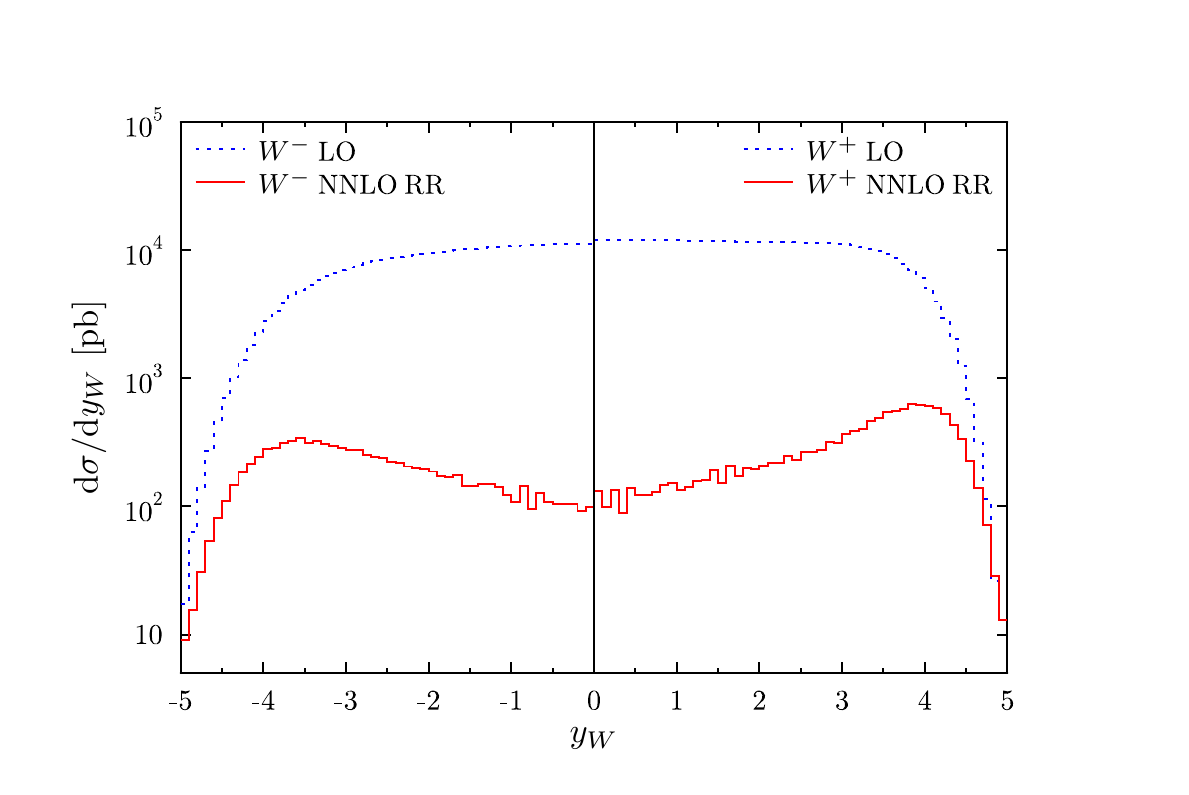}
\caption{{\label{fig:wrap}}Rapidity distribution of $W^-$ (left side) and $W^+$ (right side) at LO (blue dotted) and
the regulated double-real contribution (red) to the NNLO result.}
\end{figure}

\begin{figure}[!t]
\centering
\includegraphics[width=12cm]{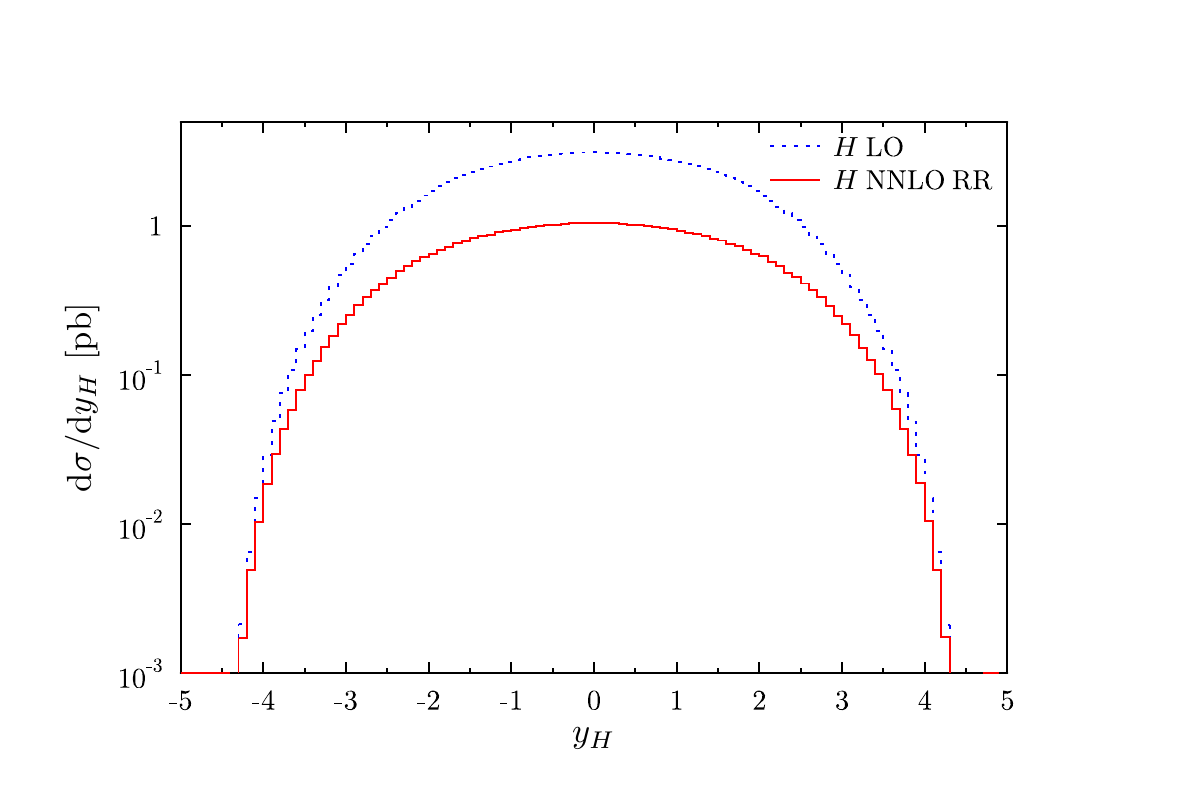}
\caption{{\label{fig:hrap}}Same as Fig. \ref{fig:wrap} but for Higgs-boson production.}
\end{figure}

With the double-real contribution regulated by subtractions it is possible to calculate the contribution
to a physical observable from that part. One natural observable for both processes is the rapidity of the
vector/scalar boson. The rapidity distribution for $W^\pm$ is shown on Fig. \ref{fig:wrap} while the 
Higgs-boson rapidity is depicted on Fig. \ref{fig:hrap}. In both cases the relative smoothness of the 
double-real contribution shows that the \colorful\ subtraction method can produce
numerically stable predictions for the most computationally demanding part.

\section{Conclusions}

In my talk I gave a status report on the application of the \colorful\ subtraction method to 
key LHC processes. The method is implemented in the \texttt{MCCSM} numerical code,
which has gone through a major revision and now includes facilities to use any PDF provider and makes 
possible to use an arbitrary number of different scales in a single 
run. This can be extremely beneficial when dealing with hadron collisions where it is not at all trivial 
how to choose the best scale for a process \emph{a priori}. As a demonstration of the method and of the 
code I presented spike and saturation plots for the double-real contribution to $W^\pm$ production and
Higgs-boson production in gluon fusion demonstrating that the newly defined subtraction terms are proper regulators. 
I also showed predictions for the computationally most demanding double-real contribution to indicate that numerically stable
predictions for differential quantities can be achieved.

\end{document}